\documentclass[prd,nofootinbib,preprint,epsfig]{revtex4}
\usepackage{latexsym} 
\usepackage{epsf}
\usepackage{epsfig}
\def\lsim{\mathrel{\rlap{\lower4pt\hbox{\hskip1pt$\sim$}}
    \raise1pt\hbox{$<$}}}      
\def\gsim{\mathrel{\rlap{\lower4pt\hbox{\hskip1pt$\sim$}}
    \raise1pt\hbox{$>$}}}      

\begin{document}
\title{Invisible Z decay width bounds on active-sterile neutrino mixing in the $(3+1)$ and $(3+2)$ models }

\author{C. A. de S. Pires \footnote{E-mail: cpires@fisica.ufpb.br}}
\affiliation{Departamento de F\'{\i}sica, Universidade Federal da
Para\'{\i}ba, Caixa Postal 5008, 58059-970, Jo\~ao Pessoa - PB,
Brazil.}


\begin{abstract}
\vspace*{0.5cm}
In this work we consider the standard model extended with singlet sterile neutrinos with mass in the  eV range  and mixed with the active neutrinos. The active-sterile neutrino mixing renders  new contributions to the invisible Z decay width which, in the case of light sterile neutrinos, depends on the active-sterile mixing matrix elements only. We then  use the current experimental value of the invisible Z decay width to obtain bounds on these mixing matrix elements for both $(3+1)$ and $(3+2)$ models. 
\end{abstract}
\maketitle

%
\section{Introduction}

Right-handed neutrinos can be introduced in the standard model in the singlet form only. Consequently, they do not interact with the standard gauge bosons. Thus people usually refers to these neutrinos as sterile neutrinos. In this context  the left-handed neutrinos, those that compose the leptonic standard doublets, are called the active ones.

Sterile neutrinos with mass in the eV range\cite{models} became popular since that LSND experiment released a report claiming the  detection of  electronic anti-neutrinos  in a beam of muonic anti-neutrinos\cite{lsndreport}. When the LSND signal is explained in terms of neutrino oscillations, then at least  one sterile neutrino is required\cite{notthree}  with mass around eV  and  mixed with the active neutrinos with a mixing angle about $\sin^2(2\theta)\approx 10^{-3}$. It happens that such range of values for mass and mixing angle  are strongly disfavored by cosmological and astrophysical data\cite{astro-cosmobounds} as well as by other short-baseline (SBL) data\cite{SBLbounds,globalanalysis}. 

In order to solve this dilemma, an experiment at Fermilab,  called MiniBooNE (MB),  was projected exclusively to confirm or refute LSND signal. Recently MB collaboration released a report which refutes the LSND signal  with 98\% CL\cite{miniboonereport}. However, according to the first global analysis after MB report\cite{firstglobalanalysis}, what is, in fact, being refuted by MB data is the possibility of explaining  the LSND signal through neutrino oscillation in the context of one sterile neutrino. Moreover they showed that if CP phase is included in scenarios with two or three sterile neutrinos then MB results can be conciliated with LSND appearance\cite{firstglobalanalysis,CPphases}. 

On the other hand, we know that  the mixing of active neutrinos with the sterile neutrinos  gives rise to interactions of these neutrinos with the standard neutral gauge bosons Z. Consequently, they will contribute to the invisible Z decay ($\Gamma_{\mbox{inv}}$). In the case of light sterile neutrinos, the $\Gamma_{\mbox{inv}}$  will present dependence on the active-sterile neutrino mixing only. The proposal of this work is to get bounds on the active-sterile neutrino mixing for both $(3+1)$  and $(3+2)$ models by using the current experimental value of the invisible Z decay width\cite{peres}.

This work is organized as follows. In Sec.~(\ref{sec2}) we settle the framework of our approach in the case of three singlet sterile neutrinos. Next, in 
Sec.~(\ref{sec3}) we obtain the bounds for the case of one sterile neutrino and  in Sec.~(\ref{sec4}) we do the same for the case of two sterile neutrinos. In   Sec.~(\ref{sec5}) we summarize our results.
\section{The $(3+3)$ model}
\label{sec2}
We develop the framework for the calculation of the $\Gamma_{\mbox{inv}}$ for the case of three singlet sterile neutrinos   added  to the standard model\cite{pleitez}. We refer to these neutrinos as  $\nu_{s_{1R}}$, $\nu_{s_{2R}}$ and   $\nu_{s_{2R}}$. The case of interest arises when we allow these singlet neutrinos to develop mixing with the active neutrinos. Such mixing  is generated by mass terms\cite{models}. The neutrino mixing matrix  $U$ in this  case is of dimension $6\times 6$ and relates the flavor eigenstates, which we consider in the base $\left( \nu_{l_L}\,,\,\nu^C_{sl_L}\right)=\left( \nu_{e_L}\,,\,\nu_{\mu_L}\,,\,\nu_{\tau_L}\,,\,\nu^C_{s_{1L}}\,,\,\nu^C_{s_{2L}}\,,\,\nu^C_{s_{3L}} \right)$, with the mass eigenstates, which we consider in the base  $\left( \nu_{L}\,,\,\nu_{sL}\right)=\left( \nu_{1_L}\,,\,\nu_{2_L}\,,\,\nu_{3_L}\,,\,\nu_{4_L}\,,\,\nu_{5_L}\,,\,\nu_{6_L} \right)$. The relation among these bases is given by 
\begin{eqnarray}
	 \left (
\begin{array}{c}
\nu_{l_L} \\
\nu^C_{sl_L}
\end{array}
\right )=U_{6\times6}
\left (
\begin{array}{c}
\nu_{L} \\
\nu_{sL} 
\end{array}
\right ),
\label{mixing}
\end{eqnarray}
 The neutrinos $\nu_1$, $\nu_2$  and $\nu_3$ are the  physical active neutrinos while  $\nu_4
$, $\nu_5$ and $\nu_6$ are the physical sterile neutrinos. In this work we neglect CP phases, which means   $U_{6\times6} $ is a real mixing matrix. 

The mixing in Eq. (\ref{mixing}) automatically generates  interactions involving the standard gauge bosons and sterile neutrinos. Here we are interested only in interactions of these light sterile neutrinos with the neutral gauge boson Z. 
 Below we present the Lagrangian that describes such interactions for the case of three sterile neutrinos
\begin{eqnarray}
	{\cal L}^\nu_Z=\frac{g}{2c_W}\left( \bar \nu_{j_L}\gamma^\mu \nu_{j_L}+ \bar \nu_{j_L}\gamma^\mu U_{ji}U_{ia} \nu_{a_L}+\bar \nu_{a_L}\gamma^\mu U_{ai}U_{i\alpha}\nu_{\alpha_L} \right)Z_\mu,
	\label{neutralcurrent}
\end{eqnarray}
where $i,j=1,2,3$, $\alpha=1,2,3,4,5,6$  and $a=4,5,6$. The Lagrangian for the case of one or two sterile neutrinos is obtained from Eq. (\ref{neutralcurrent}) by taking the corresponding values of $\alpha$  and $a$.

With the interactions given in  Eq. (\ref{neutralcurrent}), we obtain the following  expression for $\Gamma_{\mbox{inv}}$ for the case of three active neutrinos and three sterile neutrinos,
\begin{eqnarray}
\Gamma_{\mbox{inv}}=\frac{G_F m^3_Z}{4\sqrt{2}\pi}\left\{1+ \frac{1}{3}\sum_{a=4}^{6}\left[ \sum_{j=1}^{3}\left( U_{ji}U_{ia} \right)^2 +\sum_{\alpha=1}^{6}\left( U_{ai}U_{i\alpha} \right)^2 \right] \right\}.
\label{generalinvdecayformula}	
\end{eqnarray}
Note that the invisible Z decay can constrain exclusively the active-sterile neutrino mixing. In the next sections we use this expression in both $(3+1)$ and $(3+2)$ models to extract bounds on such  mixing matrix elements.

\section{The $(3+1)$ model}
\label{sec3}
The   $(3+1)$ model is the simplest sterile neutrino model. In it the mixing matrix $U$ is of dimension $4\times 4$ which, in the case of CP invariance, can be parameterized by six independent free parameters. Our considerations on these free parameters are the followings. Only two of them are really known, which are the  angles, $\theta_{23}$ and $\theta_{12}$, involved in the atmospheric and solar neutrino oscillation, respectively. The current best fit values  for these angles are $\theta_{23}=45^o$ and  $\theta_{12}=34^o$\cite{bestfit}. The third parameter we consider is the angle $\theta_{13}$. Direct searches at reactor experiments give the upper bounds $\theta_{13}\leq 12^o$, but its best fit value is $\theta_{13}=0$\cite{bestfit}. The other three free parameters are responsible for the mixing among  active and  sterile neutrinos. It is expected that these free parameters be small such that their effects on $\theta_{12}$, $\theta_{23}$ and $\theta_{13}$ could be neglected. Moreover we will make use of the fact that, in the SBL experiments, the relevant mixing matrix elements are $U_{14}$  and $U_{24}$. For example, the effective angle probed by MB and LSND experiments is $\sin^2(2\theta)=4U^2_{14}U^2_{24}$\cite{SBLbounds,globalanalysis}. Thus we will use a   $U_{4\times 4}$ Pontecorvo-Maki-Nakagawa-Sakata mixing matrix  whose parameterization focus exclusively on these mixing matrix elements. One parameterization of interest for us is given by\cite{parametrization}
\begin{eqnarray}
U_{4 \times 4}\approx	\left[ \begin {array}{cccc} c&s& 0&\delta\\\noalign{\medskip}-\frac{s}{\sqrt{2}} &\frac{c}{\sqrt{2}}&\frac{1}{\sqrt {2}}&\kappa
\\\noalign{\medskip}-\frac{s}{\sqrt{2}}&\frac{c}{\sqrt{2}}&-\frac{1}{\sqrt {2}}&
\kappa\\\noalign{\medskip}-\delta c+\sqrt {2}\kappa s&
-\delta s-\sqrt {2}\kappa c& 0&1\end {array} \right],
\label{3+1parametrization}
\end{eqnarray}
with $c=\cos\theta_{12}$  and $s=\sin\theta_{12}$. For $\theta_{12}=34^o$ we have    $c=0.83$ and $s=0.56$. Throughout this paper we use these values for $c$  and $s$.  This parameterization is interesting for our proposal because the bounds on $\delta$  and $\kappa$ fall directly on $U_{14}$  and $U_{24}$.

After these considerations we are ready to extract the bounds that $\Gamma_{\mbox{inv}}$ put on $\delta$ and $\kappa$.  For this, we  substitute the elements of $U_{4\times 4}$ given above in the expression for $\Gamma_{\mbox{inv}}$  given in Eq. (\ref{generalinvdecayformula}) for the particular case of one sterile neutrino. The current experimental value for the invisible Z decay width is  $\Gamma^{\mbox{exp}}_{\mbox{inv}}=499\pm 1.5$MeV\cite{pdg}. In this work we use  $m_Z=91.1875$GeV and $G_F=1.16637\times10^{-5}$GeV$^{-2}$.

The bounds on $\delta$  and $\kappa$ with 95\% CL are showed in FIG. 1. As we can see in that picture,  the maximum value $\delta$ and $\kappa$ can develop is $0.116$ and $0.08$, respectively. This is a very restrictive bound, which  get clear when we translate it to the effective angle that arises in appearance experiments, $\sin^2(2\theta)=4\delta^2 \kappa^2$. According to  FIG. 1, the upper bound required by the invisible Z decay width on this effective angle is of order of $\sin^2(2\theta)\leq 10^{-5}$. 

Let us confront this bound with the LSND data.  Remember that  LSND signal requires  $\sin^2(2\theta)\approx 10^{-3}$. Such value is two order of magnitude above the upper bound on this effective angle coming from the invisible Z decay width. Thus the invisible Z decay width bound on the effective angle $\theta$ enter in conflict with the possibility that LSND signal be explained by neutrino oscillation in the context of one sterile neutrino.
\section{The $(3+2)$ model}
\label{sec4}
In considering CP invariance the $U_{5 \times 5}$ mixing matrix can be parameterized by ten free parameters. Here also we consider that three of them are the angles $\theta_{12}$, $\theta_{23}$ and $\theta_{13}$ whose  current best fit values are shown in the previous section. The  other seven free parameters are responsibles by the active-sterile neutrino mixing. As in the previous case, we  consider  that these free parameters  are  small such that their effects on the angles $\theta_{12}$, $\theta_{23}$ and $\theta_{13}$ are neglected. Moreover we consider the fact that  the expression for the relevant appearance probability in SBL experiments  involve the elements $U_{14}$, $U_{15}$, $U_{24}$ and $U_{25}$ only\cite{CPphases,firstglobalanalysis,3+2globalanalysis}. Thus, following the arguments of the previous section,  the simplest  parameterization for the  $U_{5\times 5}$ Pontecorvo-Maki-Nakagawa-Sakata mixing matrix  of interest for us here is given by\cite{parametrization}
\begin{eqnarray}
U_{5 \times 5}\approx \left[ \begin {array}{ccccc} c&s& 0&\delta &\epsilon\\\noalign{\medskip}-\frac{s}{\sqrt{2}} &\frac{c}{\sqrt{2}}&\frac{1}{\sqrt {2}}&\kappa & \xi
\\\noalign{\medskip}-\frac{s}{\sqrt{2}}&\frac{c}{\sqrt{2}}&-\frac{1}{\sqrt {2}}&
\kappa &
\xi\\\noalign{\medskip}-\delta c+\sqrt {2}\kappa s&
-\delta s-\sqrt {2}\kappa c& 0&1&0
\\\noalign{\medskip}-\epsilon c+\sqrt {2}\xi s&
-\epsilon s-\sqrt {2}\xi c& 0&0&1\end {array} \right].
\label{3+2parametrization}	
\end{eqnarray}
With this mixing matrix in hand, what we have to do now is to substitute its elements in the expression for $\Gamma_{\mbox{inv}}$ given in Eq. (\ref{generalinvdecayformula}) for the case of two sterile neutrinos.  We proceed by attributing some specific values for $\xi$ and then  varying $\delta$, $\kappa$  and $\epsilon$.  We would like to stress that $\xi=0.08$ is the maximal value  that $\Gamma^{\mbox{exp}}_{\mbox{inv}}$ allows this parameter develop with 95 \% CL. The graphics in  FIG. 2 show the allowed values  $\delta$, $\kappa$ and $\epsilon$ can develop for four different values of the parameter $\xi$. As we can see in these figures, the maximum value $\kappa$  and $\epsilon$ can develop is about $10^{-2}$ while $\delta$ can attain $10^{-1}$ as maximum value. 

Just for effect of comparison, let us confront our bounds with the best fit points for these mixing matrix elements presented in the global analysis of the Refs.\cite{firstglobalanalysis,CPphases,3+2globalanalysis}. The global analysis in Refs. \cite{3+2globalanalysis,CPphases} present best fit points for the elements $U_{14}$, $U_{24}$, $U_{15}$ and $U_{25}$ while Ref. \cite{firstglobalanalysis}  presents  best fit points for the  combinations $|U_{14}U_{24}|$ and  $|U_{15}U_{25}|$. As we can see in such references, the order of magnitudes of such best fit points  lie in the range $10^{-1}$ to $10^{-2}$. Looking at the graphics displayed in FIG. 2,  we see that the values $\delta$,  $\kappa$, $\xi$   and $\epsilon$ can develop is in complete disagreement with such best fit points. This indicate that, even with two sterile neutrinos, the invisible Z decay width enters in conflict with the possibility of explaining LSND signal through neutrino oscillation. 
\section{Summary}
\label{sec5}
To summarize, we have considered scenarios with one and two light sterile neutrinos and obtained bounds on the active-sterile neutrino mixing elements from  the  invisible Z decay width. For the case of one sterile neutrino, the bound is such that the maximum values $\delta$  and $\kappa$ can develop are $0.116$ and $0.08$, respectively. This translates in the following bound on the effective mixing angle that arises in appearance experiments, $\sin^2(2\theta)=4\delta^2 \kappa^2\leq 10^{-5}$. For the case of two sterile neutrinos, the bounds on the mixing matrix elements $\delta$, $\kappa$, $\epsilon$ for four different values of $\xi$ are displayed in FIG. 2. The bounds are such that the maximum values  $\kappa$, $\epsilon$ and $\xi$ can develop are about $10^{-2}$  while $\delta$ can attain $10^{-1}$. To finalize, in both $(3+1)$  and $(3+2)$ models, the invisible Z decay width bounds on the active-sterile mixing enter in conflict with the possibility of explaining LSND signal through neutrino oscillation.

\acknowledgments
This work was supported by Conselho Nacional de Pesquisa e
Desenvolvimento(CNPq).


%
\begin{figure}
\centering
\epsfig{file=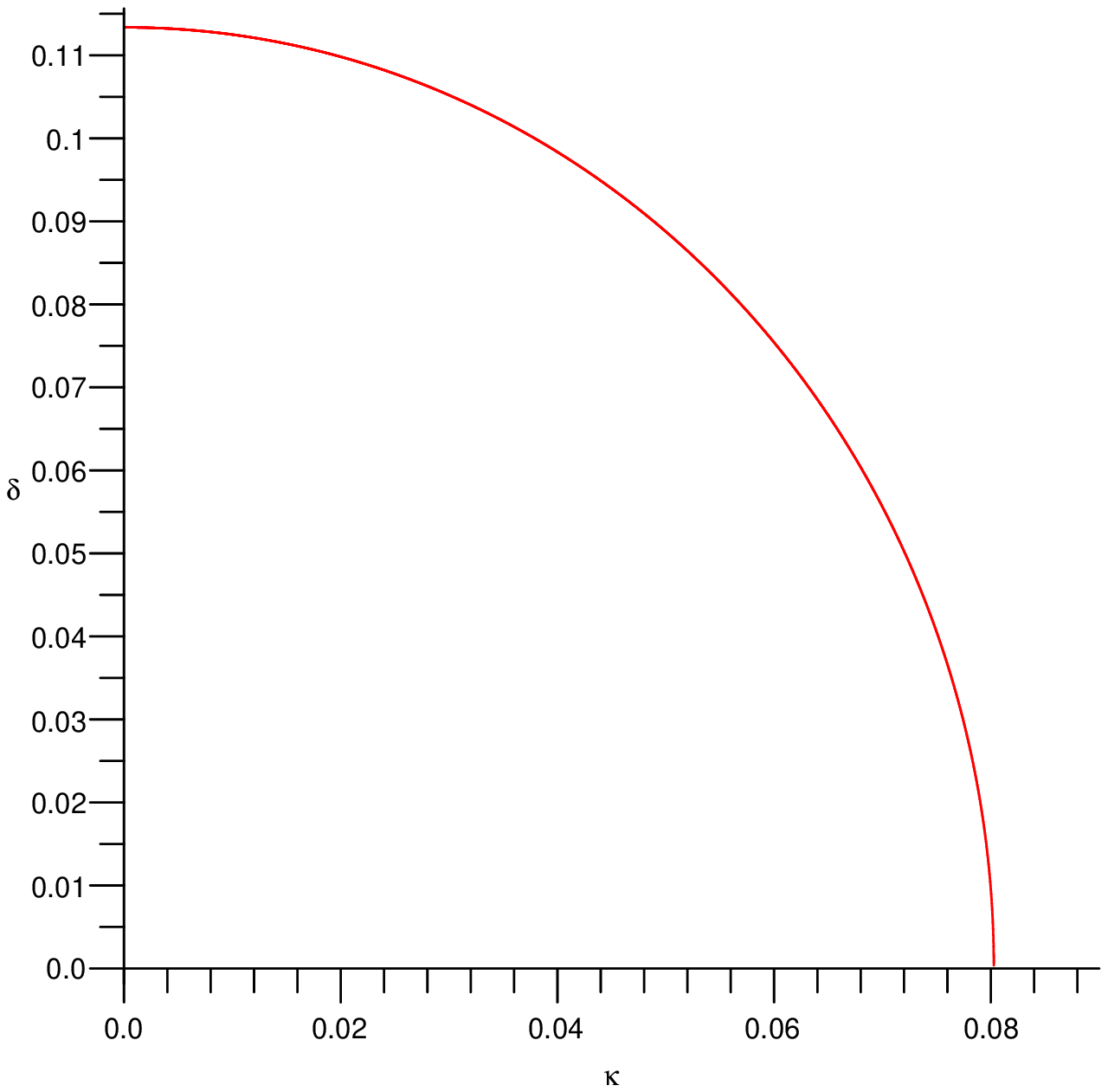,width=10cm,height=10cm,angle=0}
\caption{Possible values  that $\delta$  and $\kappa$ can develop allowed by the invisible Z decay width with 95\% CL in the ($3+1$)  model.}
\label{plot31}
\end{figure}   
\begin{figure}
\centering
\epsfig{file=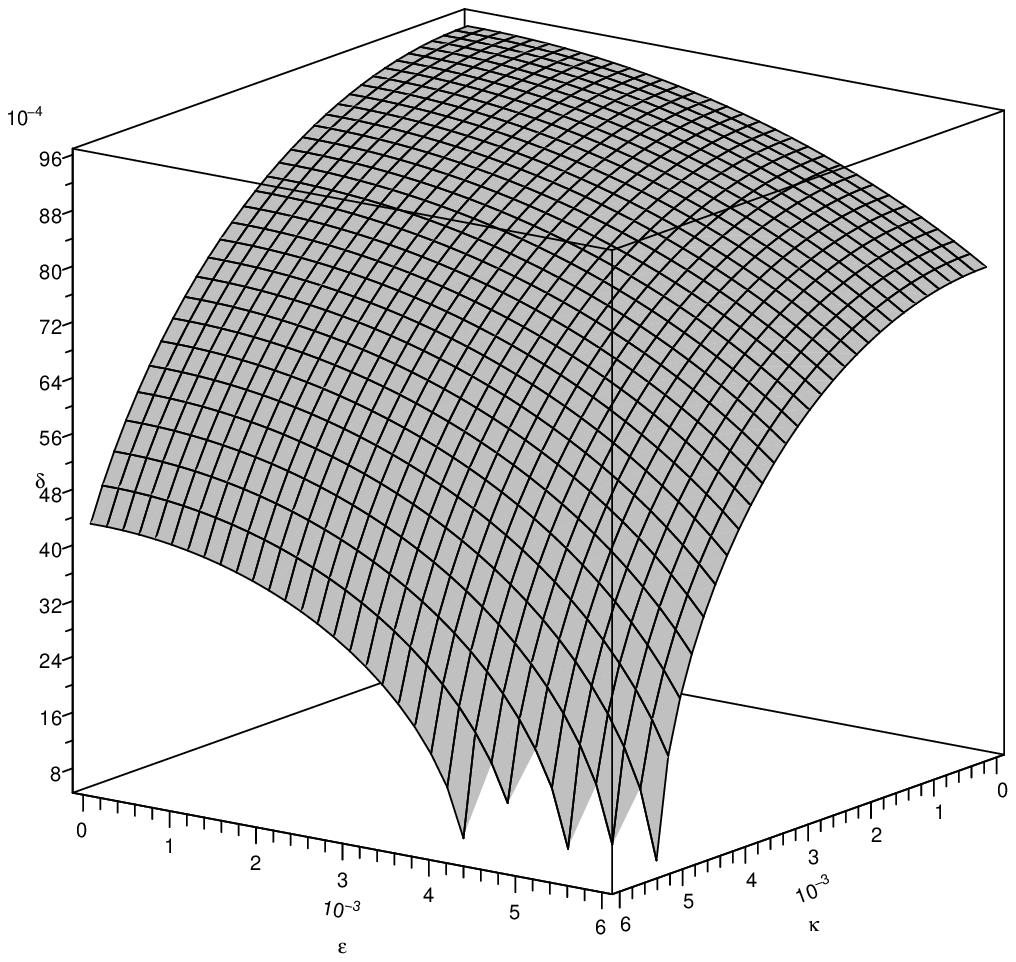,width=8cm,height=6cm,angle=0}\epsfig{file=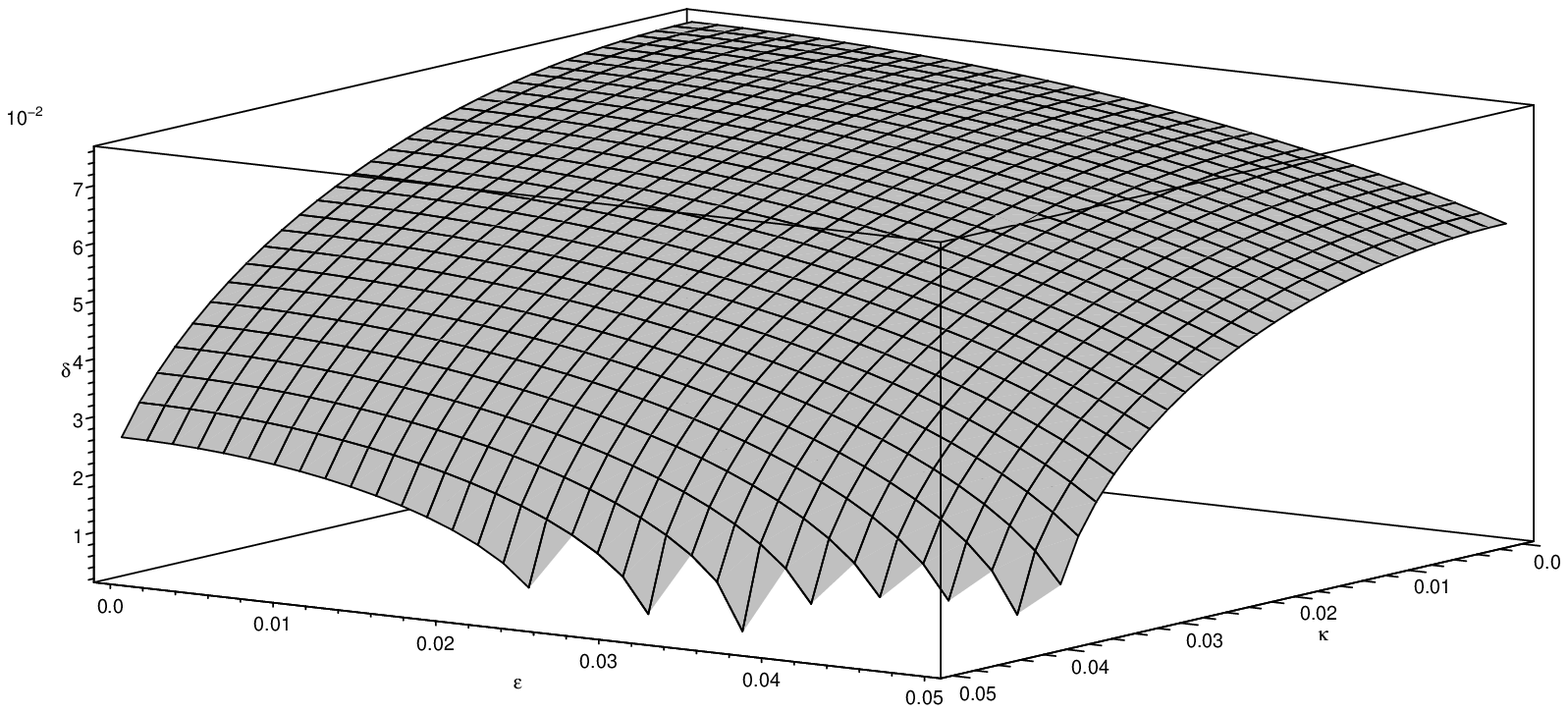,width=5cm,height=6.5cm,angle=0}\\
\,\,\,\,\,\,\,\,\,\epsfig{file=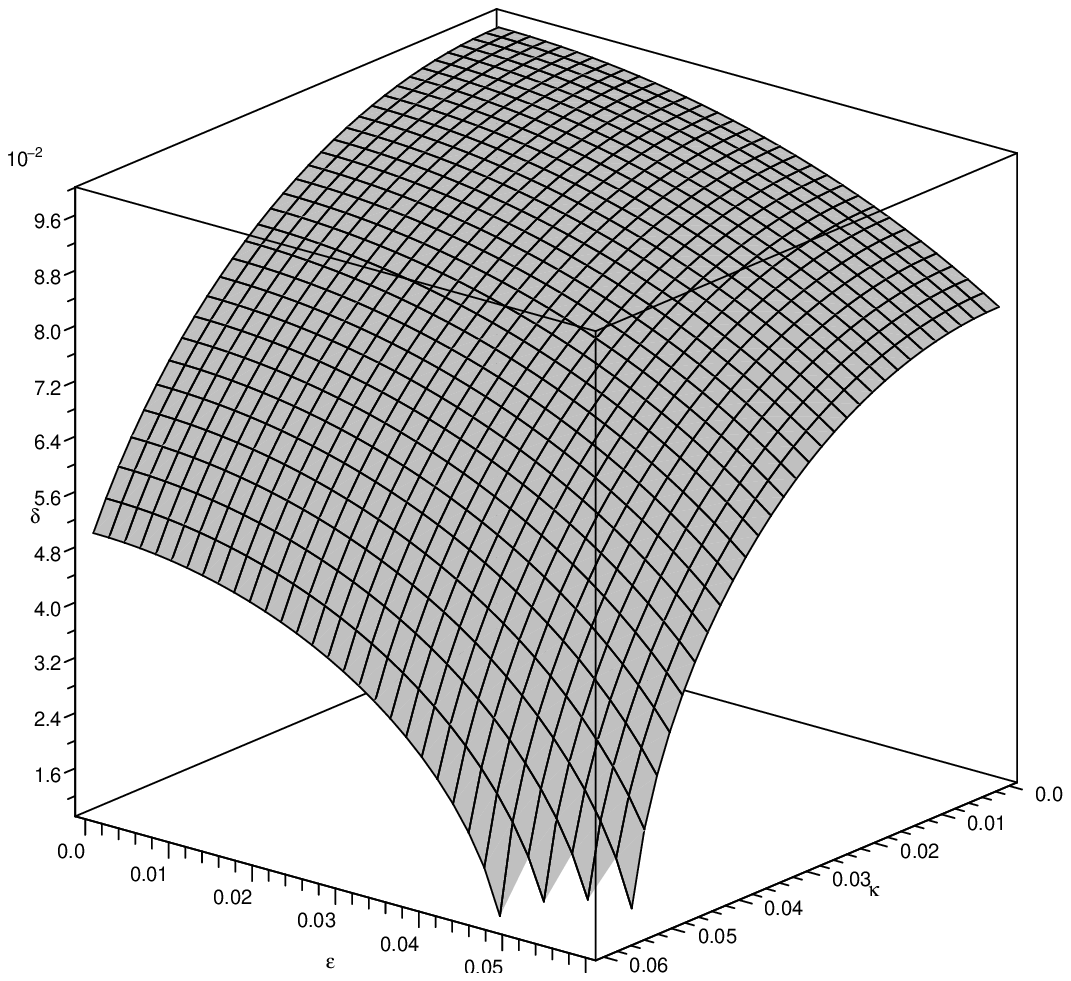,width=8cm,height=6cm,angle=0}\epsfig{file=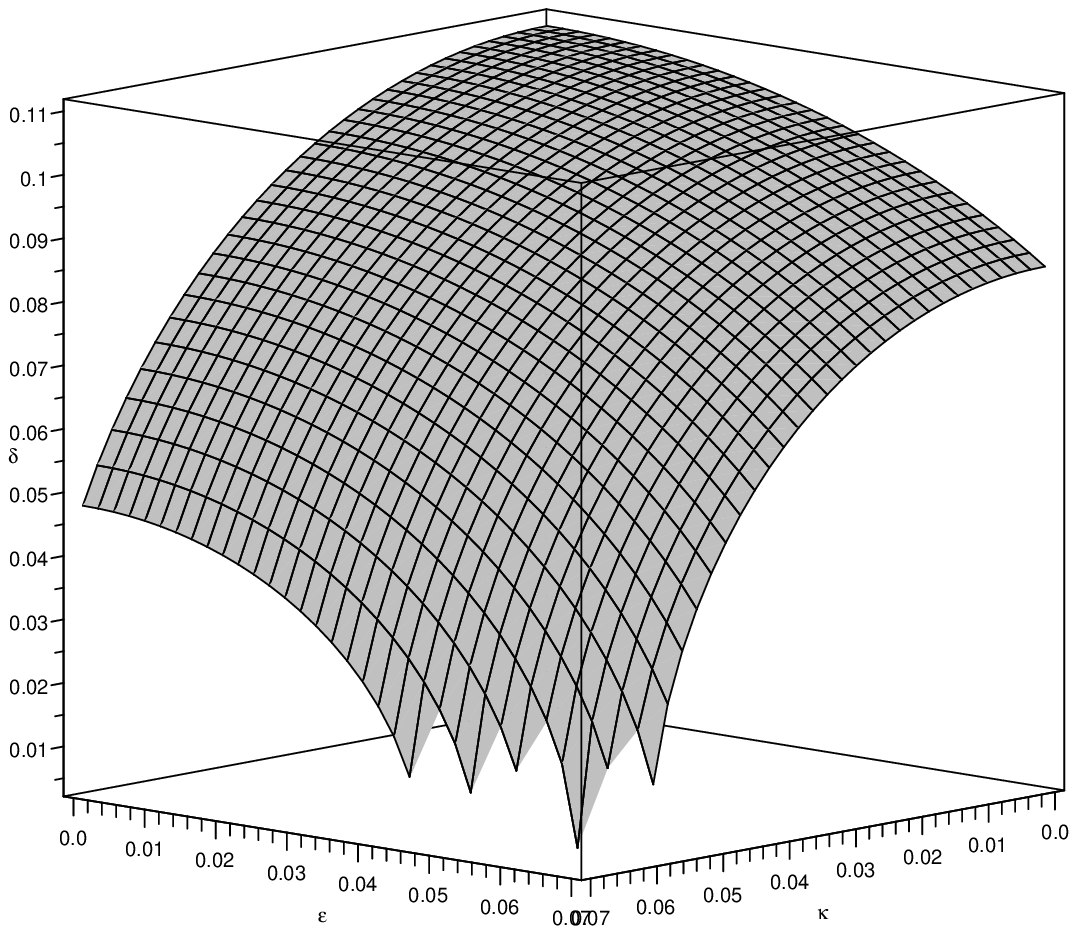,width=8cm,height=6cm,angle=0}
\caption{Possible values that $\delta$, $\kappa$ and  $\epsilon$ can develop allowed by the invisible Z decay width with 95\% CL in the ($3+2$) model for the following values of $\xi$: $\xi=0.08$(upper left), $\xi=0.06$(upper right), $\xi=0.04$(lower left), $\xi=0.02$(lower right). }
\label{plot32}
\end{figure}   
\end{document}